\documentstyle[version2,aps,eqsecnum,preprint]{revtex}
\begin{document}
\draft
\def\ltsim{\vbox {\hbox{\lower 0.9\baselineskip \hbox{$<$}} \break
                 \hbox{\lower 0.2\baselineskip \hbox{$\sim$}} } }

\begin{title}
Low Temperature Properties of Anisotropic Superconductors
with Kondo Impurities
\end{title}

\author{L.S. Borkowski and P.J. Hirschfeld}

\begin{instit}
Department of Physics, University of Florida, Gainesville, FL 32611
\end{instit}

\begin{abstract}

We present a self-consistent theory of superconductors
in the presence of Kondo impurities, using large-$N$
slave-boson methods to treat the impurity dynamics.
The technique is tested on the s-wave case and shown
to give good results compared to other methods for $T_K > T_c$.
We calculate low temperature thermodynamic and transport
properties for various superconducting states, including
isotropic s-wave and representative anisotropic model
states with line and point nodes on the Fermi surface.

\end{abstract}

\pacs{PACS numbers: 74.62Dh, 75.20Hr, 74.25Bt}

\narrowtext

\section{Introduction}

The problem of a magnetic impurity in a superconductor has been extensively
studied, but
is not completely solved because of the difficulty of treating
the dynamical correlations of the coupled impurity-conduction electron
system together with
pair correlations.  Generally,
the behavior of the system can be characterized
by the
ratio of the Kondo energy scale in the normal metal to the
superconducting transition temperature, $T_K/T_c$.
For $T_K/T_c \ll 1$,
conduction electrons scatter from classical spins and physics in
this regime can be described by the Abrikosov-Gor'kov
theory.\cite{AG} In the
opposite limit, $T_K/T_c \gg 1$, the impurity spin is screened and
conduction electrons undergo only potential scattering. In this
regime s-wave superconductors are largely unaffected by
the presence of Kondo impurities due to Anderson's theorem.\cite{anderson}
Superconductors with an anisotropic order parameter, e.g. p-wave, d-wave
etc., are strongly affected, however and the potential scattering is
pair-breaking.  The effect of pair breaking
is maximal in s-wave superconductors in the intermediate region,
$T_K \sim T_c$, while in the anisotropic case it
is largest for $T_K/T_c \rightarrow \infty$.\cite{lbspjh}

Regardless of the type of the pairing mechanism, the conduction band
density of states is depleted in the neighborhood of the Fermi surface,
resulting in the weakening of the coupling to the impurity. On the
other hand, increasing the impurity concentration or the coupling
strength reduces $T_c$.  This competition is influenced strongly
by the number of states removed from near the
Fermi surface by the formation of the superconducting state.  In fact,
Withoff and Fradkin\cite{wf} have pointed out that for the
normal state analogue of this problem, in which the conduction band
is assumed to behave as a semimetal, with density of states varying as
$N(\omega)\simeq |\omega|^r$, $r>0$,
there is a critical coupling strength $J_c$ below which the Kondo effect
no longer  occurs.
One might expect the consequences of such a transition, were it to take
place, e.g., as a function of temperature in an unconventional
superconducting state providing the power-law density of conduction electron
states, to be dramatic. In a previous study of this problem, however,
we have shown that under most circumstances
a finite density of Kondo impurities in the thermodynamic limit induces
a finite density of states at the Fermi level, ensuring that a Kondo
effect always occurs, but with reduced effective Kondo temperature.
Nevertheless, unusual pairbreaking effects
are still to be expected in unconventional superconductors due to the dynamical
reduction of the Kondo scale caused by pair correlations.

While this work was begun in anticipation of applications to heavy
fermion superconductors, recent measurements of penetration
depth,\cite{Hardypendepth}
photoemission,\cite{photo} and Josephson tunneling
on YBCO and BSSCO\cite{squid} have provided evidence that the copper
oxide superconductors may be unconventional, possibly d-wave as well.
This conclusion remains controversial, however, and experimental tests to
distinguish conventional from unconventional pairing are of great current
interest.  One would like, for example, to develop an understanding
of the effect of doping Zn and Ni impurities in
the ${\rm CuO_2}$ planes. Simple
models of Zn and Ni acting as strong and weak potential scatterers in a d-wave
superconductor, respectively, are consistent with some experiments at low
doping levels,\cite{NMR,Bonnhardybigpaper,Walstedt} but
inconsistent with other
measurements.\cite{Ulmetal,Alloul}
Since in some cases Zn impurities appear to possess a magnetic moment
at higher temperatures,\cite{Alloul} it is of interest to explore
whether an s-d type exchange coupling of conduction electrons
to an impurity embedded in an unconventional
superconductor can describe the range of behavior observed.

In this work we focus on basic thermodynamic and transport properties
of superconductors doped with Kondo impurities.
Our aim is to develop a tractable, self-consistent scheme for the
calculation of all basic properties of superconductors using methods known
to successfully describe the most difficult aspect of the problem, namely
the dynamics of the Kondo impurity.  For this reason we have adopted the
large-$N$ "slave boson" approach of Barnes,\cite{barnes}
Coleman,\cite{coleman} and Read and Newns,\cite{rn} as this approach
is well-known to provide a good description of the spectral properties of
the Kondo impurity for sufficiently low temperatures.  We emphasize that
this description is adequate for most of our purposes ($T_c \ltsim T_K$)
even in the case
of spin degeneracy $N=2$, since the impurity spectral resonance is located
at the proper position, i.e. exactly at the Fermi level in this approximation.

\section {Formalism}

We use the large-$N$ slave boson technique for the $SU(N)$ Anderson
model describing an $N$-fold degenerate band of conduction
electrons, $c_{km}$, $m=1,...N$ with
energy $\epsilon_k$ hybridizing through matrix element $V$ with a localized
impurity state $f_m$.
In general this Hamiltonian contains a term with the
Coulomb repulsion $U$ between two electrons present at the impurity
simultaneously. In many compounds $U$ is large (5--10 $eV$ in the
lanthanides) and for the purpose of studying the low temperature physics
we will assume $U=\infty$. The on-site repulsion term  is then
absent, but a constraint is added to
ensure that the system remains in the physical part of the Hilbert
space. The conduction band is assumed to have a constant density of
states in its normal state, $N(\omega)=1/2D=N_0$. We also include
a BCS-like pairing term of
electrons on opposite sides of the Fermi sphere,

\begin{eqnarray}
H & = & \sum_{k,m} \epsilon_k {c^{\dagger}}_{km} c_{km} + E_f
\sum_m {f^{\dagger}}_m f_m + V \sum_{k,m} [{c^{\dagger}}_{km}f_m b
+ h.c.]
 \nonumber \\
& + & \sum_{k,m}[\Delta(k){c^{\dagger}}_{km}{c^{\dagger}}_{-k-m} + h.c.]
+ \lambda (\sum_{m} {f^{\dagger}}_m f_m + b^{\dagger}b) ,
\end{eqnarray}
\noindent
where $\lambda$ is a Lagrange multiplier enforcing the constraint
$n_f+n_b = \sum_{m} {f^\dagger}_m f_m + b^\dagger b = 1$,
preventing double occupancy of the impurity
site.  In the limit $E_f \rightarrow -\infty$, $N_0 V^2/E_f = $const,
Eq. (2.1) reduces to the Coqblin-Schrieffer Hamiltonian with pairing
studied in Ref. \cite{lbspjh}.  Here we have chosen the more general
form (2.1) to study deviations from single occupancy, $n_f \ne 1$,
although we do not attempt to explore the fully developed
mixed valent regime.

The mean field approximation to this model, with mean-field
amplitude $\langle b \rangle$, leads to the two equations

\begin{equation}
{1\over N} = - {\rm Im} \int_{-\infty}^\infty d\omega f(\omega)
{1\over 2} {\rm Tr}(\tau_0 + \tau_3) {\bf G}_f (\omega+i0^+) ,
\end{equation}
\noindent
and

\begin{equation}
{{E_f-\epsilon_f}\over V^2}={\rm Im}\int_{-\infty}^\infty d\omega f(\omega)
{1\over 2}{\rm Tr} \left[({\tau_0+\tau_3})({\bf G}^0(k,\omega+i0^+)
{\bf G}_f(\omega+i0^+))\right] ,
\end{equation}
\noindent
which determine $\langle b \rangle$ and $\epsilon_f$,
the latter being the position of the resonant
state. Eqs. (2.2) and (2.3) should be solved self-consistently
together with the gap equation,

\begin{equation}
\Delta(k) = -T \sum_{\omega_n} \sum_{k^\prime} V_{kk^\prime}
{\rm Tr} {\tau_1\over 2} {\bf G}(k^\prime,\omega_n).
\end{equation}
The full conduction electron Green's function ${\bf G}$ and
the impurity Green's function ${\bf G}_f$
\begin{equation}
{\bf G}=\left(\matrix{G&F\cr
		     F^\dagger&G^*}\right)
\quad {\rm and} \quad
{\bf G}_f=\left(\matrix{G_f&F_f\cr
		     {F_f}^\dagger&{G_f}^*}\right)
\end{equation}

\noindent
are calculated from the diagrams shown in Figure 1, yielding
\begin{equation}
{{\bf G}(\omega)}^{-1}={{\bf G}^0(\omega)}^{-1} - {\bf \Sigma}(\omega)
= {\tilde \omega} \tau_0 - \epsilon_k \tau_3 - {\tilde \Delta(k)} \tau_1 ,
\end{equation}
\begin{equation}
{{\bf G}_f(\omega)}^{-1} = {{\bf G}^0_f(\omega)}^{-1}
- {\bf \Sigma}_f(\omega)
={\bar \omega} \tau_0 - \epsilon_f \tau_3 - {\bar \Delta} \tau_1 .
\end{equation}

\noindent
The Green's functions are now averaged over impurity positions
in the usual way. The renormalized frequencies are calculated
self-consistently from the Dyson equations,

\begin{equation}
{\tilde \omega} = \omega + \alpha {\bar \omega}
/(-{\bar \omega}^2 + {\epsilon^2}_f+{\bar \Delta}^2) ,
\end{equation}
\noindent
and

\begin{equation}
{\bar \omega} = \omega + \Gamma {\left \langle {\tilde \omega}
/{\left ( {\tilde \Delta}^2(k) - {\tilde \omega}^2 \right )}^{1/2}
\right \rangle}_{\hat k} .
\end{equation}
\noindent
Here $\alpha = N {\bar n}\Gamma T_{c0}/2\pi$ ,
${\bar n} = n/N_0T_{c0}$
is the scaled impurity concentration,
and $\langle ... \rangle _{\hat k}$ is
a Fermi surface average. The off-diagonal
renormalizations are

\begin{equation}
{\tilde \Delta}(k) = \Delta(k) + \alpha {\bar \Delta}
/(-{\bar \omega}^2 + {\epsilon^2}_f + {\bar \Delta}^2)  ,
\end{equation}
and

\begin{equation}
{\bar \Delta} = \Gamma {\left \langle {\tilde \Delta}(k)
/{\left ( {{\tilde \Delta^2(k)} - {\tilde \omega}^2} \right )}^{1/2}
\right \rangle}_{\hat k}  .
\end{equation}
In superconductors with order parameters where the Fermi surface average in
Eq. (8) vanishes,
off-diagonal corrections vanish and ${\tilde \Delta}(k) = \Delta (k)$.
This class includes but is not limited to odd-parity superconducting
states.

The energy scale $\Gamma$ is the renormalized resonance width,
$\Gamma = {\langle b \rangle}\pi N_0 V^2$.
The low-temperature Kondo scale in the large-$N$ slave boson theory
is given by $T_K\equiv\sqrt{\Gamma^2+{\epsilon_f}^2}$.  Although
the width of the actual spectral feature corresponding to the Abrikosov-Suhl
resonance is modified below $T_c$, in what follows
we will normalize
all quantities with respect to this $T_K$, evaluated from Eqs. (2.2)
and (2.3) with $\Delta=0$ at $T=0$.  In the regime of principal interest,
$T_K>T_c$, corrections  to this definition will be small in any case.

\section{Results for an s-wave superconductor}

The early history of the problem of a Kondo impurity
in an s-wave superconductor has been reviewed
by M\"uller-Hartmann.\cite{mh}
Abrikosov and Gor'kov first discussed the pair-breaking effects
of magnetic impurities weakly coupled via exchange interactions
to conduction electrons.\cite{AG}
Shiba\cite{shiba0} extended this
approach to treat strong scattering by classical spins,
using the $t$-matrix approximation, and showed the existence
of bound states in the gap.
At finite concentration of impurities the bound states were found to
form an impurity band whose width and center scaled with impurity concentration
and exchange strength. These early works neglected the dynamical screening
of the localized spin by the conduction electron gas.
These effects were
incorporated by M\"uller-Hartmann and Zittartz,\cite{mhz1} adopting an
equation of motion decoupling scheme  previously used by
Nagaoka\cite{nagaoka} to calculate the
dynamical spin correlations in  the normal state.
This approach correctly reproduced results in the Abrikosov-Gor'kov
limit, $T_K/T_c \rightarrow 0$,
and made the remarkable prediction  of a reentrant superconducting phase
if $T_K \ll T_c$, subsequently observed in experiments
on ${\rm La_{1-x}Ce_xAl_2}$.\cite{riblet,maple1}
The failure of the decoupling scheme used to capture the correct
crossover to Fermi liquid behavior in the normal state as $T\rightarrow 0$
invalidated the  M\"uller-Hartmann--Zittartz approach in the low
temperature regime $T_K>T_c$, however.

The physics of the Fermi liquid regime, $T_K/T_c \rightarrow \infty$,
was studied by  Matsuura, Ichinose and
Nagaoka\cite{min} and by Sakurai\cite{sakurai} by extending
the Yamada-Yosida theory\cite{yy}
to the superconducting state.
They obtained an exponential $T_c$-suppresssion with increasing
impurity concentration $\bar n$, $T_c\simeq T_{c0} \exp (-p{\bar n}/\lambda)$,
where $\lambda$ is the BCS dimensionless coupling constant, and $p$ is a
constant of order unity.
This is commonly referred to as "pair-weakening" as opposed
to pair-breaking, since the effective superconducting coupling constant
is reduced due to correlations on the impurity site.
The exponential form breaks down for concentrations
sufficiently close to a critical ${\bar n}_c$, for which $T_c=0$.
In this regime
the reentrant behavior found by  M\"uller-Hartmann and Zittartz
does not occur.  A further characteristic signature of the Fermi liquid
regime is the reduced specific heat jump, $C^*\equiv (\Delta C/\Delta C_0)
/(T_c/T_{c0}))_{|T_c = T_{c0}} $ which is always less than
one,\cite{sakurai,ichinose}
in contrast to the high temperature regime.

Not surprisingly, qualitatively similar results were obtained  by
other early workers for Kondo and Anderson impurities
using a variety of other approaches.\cite{mh}
More recent treatments include the use of
a self-consistent large-$N$\cite{bickers,lbspjh}, Monte Carlo\cite{jarrell1}
and NRG methods.\cite{shiba4}
Schlottman\cite{Schlottmann} has treated the mixed--valence
regime using Brillouin-Wigner perturbation theory.
Out of these efforts has evolved a qualitatively consistent
picture of the effect of Kondo
impurities on the superconducting transition,\cite{maple2}
but little understanding of the low-temperature properties
of Kondo-doped superconductors because of the difficulty
of the calculations involved. In this section we show that
the current theory reproduces the known effects
of Kondo impurities on the critical temperature, specific
heat jump and bound states spectrum of an s-wave superconductor.

\subsection{Critical Temperature}

The simplest and most direct  effect of impurity scattering on
a superconductor is the suppression of the critical temperature.
Scattering from impurities with internal quantum-mechanical
degrees of freedom leads to deviations from the classic Abrikosov-Gor'kov
prediction for the dependence of $T_c$ on impurity concentration.\cite{AG}
These effects depend sensitively on the low-energy behavior of the
self-energy $\Sigma(\omega)$, which enters  the linearized
gap equation, obtained from Eq. (2.4) near $T_c$,

\begin{equation}
{\rm ln} (T_c/T_{c0}) = 2\pi T_c \sum_{n \ge 0} {1\over
{ \omega_n ( 1 + \alpha/ B(\omega_n))} }
- \sum_{n \ge 0} {1\over {n+1/2}} ,
\end{equation}
where $B(\omega_n)=(\omega_n + \Gamma)^2 + {\epsilon_f}^2$ .

The slope of the $T_c$-suppression evaluated at ${\bar n=0}$ is
\begin{equation}
{\left ({1\over T_{c0}}{{dT_c}\over {d{\bar n}}} \right )}_{{\bar n}=0}
= - {N\over {2\pi}} \sum_{n \ge 0} {{\Gamma T_{c0}}\over
{(n+1/2) B(\omega_n)}} .
\end{equation}

\noindent
In an insert to Figure 2, we have plotted a numerical evaluation of Eq. (3.2)
for  an s-wave superconductor.
Note the curve is drawn with a broken line for
small $T_K/T_{c0}$ to reflect the fact that the
slave boson mean field theory breaks down there.
The maximum of the $T_c$--suppression is found to occur at $T_K \simeq 3T_c$,
similar to the NCA result.\cite{bickers}
The early high-temperature
theory of M\"uller-Hartmann and Zittartz\cite{mhz1} locates this maximum at
$T_K/T_c \simeq 12$, whereas
in the more recent Monte Carlo calculation\cite{jarrell1}
for the symmetric Anderson model with finite $U$
the maximum slope of the $T_c$-suppression is at $T_K \simeq T_c$.
Unfortunately a direct quantitative comparison with the latter work
is not possible, as the simulation is not performed in the fully
developed Kondo regime.

The present theory predicts an exponential decrease
of $T_c$
at small concentrations, in agreement with other
theories of the Fermi liquid regime,\cite{sakurai,ichinose}

\begin{equation}
T_c \simeq T_{c0} {\rm exp}{\left [-{\alpha \over {T_K}^2}
\Psi(T_K/2\pi T_{c0}) \right ]} \simeq T_c
{\left (1-{\alpha \over {T_K}^2} {\rm ln}(T_K/T_{c0}) \right ) ,}
\end{equation}
where $\Psi$ is the digamma function. In Refs. \cite{matsuura}
and \cite{sakurai} the initial suppression of $T_c$ is proportional
to ${\rm ln}^2(T_K/T_{c0})$.   A full evaluation of  Eq. (3.2) for arbitrary
concentrations and various values of the ratio $T_K/T_c$ is shown in Figure 2.
It is interesting to note that the theory reproduces the reentrant
behavior characteristic of the high temperature regime\cite{mh}
although we do not expect the theory to be accurate in this case (dashed
line).

As is evident in the Figure the current theory predicts no critical
concentration $n_c$ for which $T_c=0$.  This is a subtle point discussed
by Sakurai,\cite{sakurai} who suggests that a failure to include
the dynamics of magnetic scattering by states close to the Fermi surface
can lead to such an effect.  Such processes are included in the
finite-$U$ perturbation theory through Coulomb vertex corrections to
the impurity averaged pair correlation function.  In our $U=\infty$
theory, such vertex corrections arise
first in leading order $1/N$ corrections
due to the exchange of slave bosons, whose dynamics are neglected in
this work.  We expect that effects arising from the absence
of these fluctuations in the theory will be quantitatively small for
$T_c\ltsim T_K$, except for impurity concentrations so large such
that $T_c\ll T_{c0}$.

\subsection{Bound states}
For conventional superconductors, $\Delta(k)=\Delta$,
Eqs. (2.9) and (2.11) are simply

\begin{equation}
{\bar \omega} = \omega + \Gamma  {{\tilde \omega}\over
{\left ( {\tilde \Delta}^2 - {\tilde \omega}^2 \right )}^{1/2}} ,
\end{equation}
and

\begin{equation}
{\bar \Delta} = \Gamma {{\tilde \Delta}\over
{\left ( {{{\tilde \Delta}^2} - {\tilde \omega}^2} \right )}^{1/2}}  .
\end{equation}

\noindent
It is obvious from Eq. (3.5) that a finite gap in the conduction
electron spectrum induces a gap in the impurity spectrum. When the
Abrikosov-Suhl resonance, which develops at temperatures below
$T_K$, falls
in the superconducting gap, bound states appear in the conduction
electron spectrum.
These peaks are placed symmetrically relative to the center
of the gap and their spectral weight depends on $T_K/T_c$ and on impurity
concentration. The density of states for $N=2$ and $N=4$ is shown
in Figure 3.
In the high-temperature regime the bound states
emerge from the edges of the gap and move towards the center of the gap
as $T_K/T_c$ increases.\cite{mhz2} In the low-temperature limit,
the bound states disappear
into the gap edges again.\cite{matsuura}
Recently, Shiba et al.\cite{shiba4} studied the position of
bound states in the gap in an s-wave superconductor using the numerical
renormalization group (NRG). Their results cover both the high- and
low-temperature limits reliably.
The Monte Carlo study
by Jarrell et al.\cite{jarrell2} confirms the overall dependence
of $\omega_B$ on $T_K/T_c$.

Bound states correspond to the poles of the t-matrix for
conduction electrons, ${\bf T}(\omega) = V^2{\bf G}_f(\omega)$
and are given by

\begin{equation}
|\omega| = \left ({{T_K}^2 + {{\Gamma^2\omega^2}\over
{\Delta^2-\omega^2}}} \right )^{1/2} -
{{\Gamma|\omega|}\over {\left ({\Delta^2-\omega^2} \right )}^{1/2}}  .
\end{equation}
We compare the solution of Eq. (3.6) with the NRG result in Figure 4. There
is a good agreement for $T_K > T_c$. For $T_K \gg T_c$
the position of bound states $\omega_B/\Delta$ is a
quadratic function of $T_K/T_c$,
$|\omega_B/\Delta| \simeq 1 - 2\Delta^2/{T^2}_K$.
The discrepancy between the NRG result and our calculation for
$T_K < T_c$ is not
surprising since the slave-boson theory fails in the high
temperature regime. The slave-boson
mean-field amplitude vanishes around $T \sim T_K$
and the theory is unable to describe the crossover to temperatures above $T_K$.
In Figure 4 we also show the spectral weight of the bound states
in the gap which again agree with the NRG calculation for
$T_K \gg T_c$.\cite{shiba4}

To calculate the specific heat jump we use equations (2.5), (2.7) and (3.1),
(3.2), and expand the gap equation near $T_c$ in terms of $\Delta/T$,

\begin{equation}
{\rm ln} {T_{c0}\over T} = - \sum_{n\ge 0}
{1\over {(n+1/2)(1+\alpha/B(\omega_n))}} +
\sum_{n\ge 0} {1\over {n+1/2}} +
{1\over 2}b_1 {\left ({\Delta \over {2\pi T}} \right )}^2  ,
\end{equation}
where

\begin{equation}
b_1=\sum_{n\ge 0}
{{(1+2\alpha\Gamma\omega_n/B^2(\omega_n)E(\omega_n))}\over
{{(n+1/2)}^3 E^3(\omega_n)}}  ,
\end{equation}
and  $E(\omega_n)=1+\alpha/B(\omega_n)$.
After expanding the free energy to fourth order in $\Delta$ we obtain

\begin{equation}
C_s(T_c) - C_n(T_c) = {{8\pi^2N_0T_c}\over b_1}
{\left ( 1-\sum_{n\ge 0}{{4\pi\alpha T_c(\omega_n+\Gamma)
\over (B(\omega_n)+\alpha)^2}} \right ) }^2  .
\end{equation}

\noindent
As can be seen in Figure 5, the dependence
of $\Delta C/\Delta C_0$ on $T_c/T_{c0}$, for $T_K \ltsim T_{c0}$, is
qualitatively different from the Fermi liquid regime.
In that limit,
$C^* \simeq 1-1/{\rm ln}(T_K/2T_{c0})$, see Figure 6.
Ichinose\cite{ichinose} and Sakurai\cite{sakurai2}
have obtained a somewhat different
result, $C^* \simeq 1-1/{\rm ln}^2(T_K/T_{c0})$.

\section{Impurities in unconventional superconductors}

The problem of an Anderson impurity in an unconventional superconductor
is of considerable interest for several reasons.  First, it serves as a testing
ground for the ideas of Withoff and Fradkin, who argued that in the analogous
Kondo problem with power law conduction electron density of states,
$N(\omega) = C|\omega|^r$, there existed a critical coupling below
which impurities are effectively
decoupled from the conduction band.\cite{wf,Soda}
We showed in Ref. \cite{lbspjh} that in an
unconventional superconductor a critical coupling indeed exists for the
1-impurity problem, but that for a finite density of impurities
there is always a finite density of conduction electron states
at the Fermi level, provided $N=2$ or if $r \le 1$.
This is consistent with phenomenological studies of potential scattering
in unconventional superconductors.\cite{hvw,smv}
This does not exclude the possibility
of observing some vestige of this transition as the Kondo temperature
becomes quite small, however.

Secondly, Kondo impurities in unconventional superconductors
have been proposed as analogues of defects in Kondo lattices,
\cite{varma} with the argument that a vacancy
in a Kondo lattice may induce a
{\it relative} phase shift close to $\pi/2$.  While such a picture
clearly cannot account for the degradation of coherent transport by
such defects, it is interesting to further explore the analogy, which has been
successful in describing the superconducting state of heavy fermion
materials.

Finally, there is currently considerable interest in the possibility that
the cuprate superconductors may be characterized by an unconventional
order parameter, as suggested by both
experimental\cite{Hardypendepth,photo,squid,NMR} and
theoretical\cite{dwavetheory} studies.
Studies of ${\rm YBa_2(Cu_{1-x}M_x )_3 O_7}$, where M are Zn or Ni dopants
in the CuO planes, suggest that  the behavior of the two ions
can be quite different.
Maple and collaborators have claimed
that Pr may act as a Kondo-like dopant in YBCO.  We believe
that a Kondo- or Anderson-type
model may be sufficiently rich to describe the unusual features of the
Zn and Ni doping studies as well, in conjunction with an understanding
of the different gap nodes in unconventional superconductors.  To
this end we study the effects of Kondo impurities on model unconventional
states, bearing in mind that the theory may be inadequate to describe
the fully developed magnetic regime $T_K \ltsim T_c$.

We chose to do our calculations for the axial,
$\Delta(k)=\Delta_0 ({\hat k}_x + i {\hat k}_y)$,  and polar,
$\Delta(k)=\Delta_0 {\hat k}_z$, states
for simplicity. These two states, nominally p-wave pairing states over
a spherical Fermi surface in 3D, are quite generally representative
of two classes of order parameters. States with order
parameters vanishing at points on the Fermi surface, like the axial
state, have low temperature
properties associated with a quasiparticle density of states
$N(\omega) \sim \omega^2$,
whereas the states with lines
of nodes, like  the polar state,
correspond to $N(\omega)\sim \omega$.
More complicated order parameters, such as those having a $d$-wave symmetry,
will have similar properties at low temperatures, since
the main factor determining the low-$T$ properties is the order
parameter topology, i.e. whether there are points
or lines of zeros of the order parameter.
In the case of points (lines) of nodes,
the low-temperature specific heat of pure superconductors
is proportional to $T^2$ ($T$).
The deviation of the penetration depth $\Delta\lambda$ from
its zero temperature value $\lambda(0)$
along the main axes of symmetry of the order
parameter is either $\sim T^2$ or $\sim T^4$ in the axial state,
according to the direction of current flow,\cite{grossetal}
whereas for the polar state it is either $\sim T$ or $\sim T^3$.
The presence of strong impurity scattering complicates the picture
and these power laws do not hold in general.

Due to the absence of the Anderson theorem for p-wave-like or
d-wave-like superconductors,
the influence of impurity scattering on the critical temperature
is qualitatively different from that
of  s-wave-like superconductors.\cite{gorkov,ru}
For the unconventional states of interest,
there are no off-diagonal corrections to
the superconducting Green's function, ${\tilde {\Delta} = \Delta}$.
In this case $T_c$ is determined by
\begin{equation}
{\rm ln} (T_c/T_{c0}) = 2\pi T_c \sum_{n \ge 0} {1\over
{ \omega_n ( 1 + \alpha /B(\omega_n)) + \alpha/B(\omega_n)}}
- \sum_{n \ge 0} {1\over {n+1/2}}.
\end{equation}
The initial $T_c$-suppression is then given by
\begin{equation}
{\left ({1\over T_{c0}}{{dT_c}\over {d{\bar n}}} \right )}_{n=0}
= - {N\over {4\pi^2}} \sum_{n \ge 0} {{\Gamma (\omega_n+\Gamma)}\over
{{(n+1/2)}^2 B(\omega_n)}} ,
\end{equation}
and approaches $-N\Gamma^2/8T^2_K$, in the limit
$T_K/T_{c0} \rightarrow \infty$, see Figure 7.
Note that ${\tilde \omega}_n/\Delta \rightarrow \alpha \Gamma/\Delta T_K^2$ ,
as $T_c \rightarrow 0$ , and as a consequence the critical concentration
$n_c$ is finite.

The specific heat jump at $T_c$ can  be found by the method already mentioned
in the preceding section,

\begin{equation}
C_s(T_c)-C_n(T_c) = {{8\pi^2N_0T_c}\over b_1}
{\left [ 1-\sum_{n\ge 0}
\left ( {G(\omega_n) \over {((n+1/2)E(\omega_n)+\alpha/B(\omega_n))^2}}
\right ) \right ]}^2  ,
\end{equation}
where

\begin{equation}
b_1=\sum_{n\ge 0} {{b+cH(\omega_n)}\over
{{[(n+1/2)E(\omega_n)+\alpha/B(\omega_n)]}^3}} ,
\end{equation}
with $b=4/5$ $(3/5)$, and $c=2/3$ $(1/3)$, for the axial (polar)
state, and $H(\omega_n)=\alpha \Gamma \left [ {2(\omega_n+\Gamma)^2}
/B(\omega_n)-1 \right ]/B(\omega_n)$  ,
$G(\omega_n)=2\alpha\omega_n(\omega_n+\Gamma)(n+1/2+\Gamma/2\pi T_c)
/B^2(\omega_n) + \alpha\Gamma/2\pi T_c B(\omega_n)$ .

The effect of Kondo impurities on the density of states
is shown in Figure 8. As in the s-wave case, the resonant states move
towards the edges of the gap as $T_K/T_c$ increases, except in the
special case $N=2$, where they are pinned at the gap center.\cite{lbspjh}
In all cases the impurity bands are broadened  relative to the s-wave case
by the continuum in which they are embedded.
In the limit $T_K\rightarrow\infty$,
the current theory coincides with the results given by phenomenological
t-matrix treatments\cite{hvw,smv} for $N=2$.

To obtain the specific heat we differentiate the entropy with
respect to temperature $C=TdS/dT$. The entropy is given by

\begin{equation}
S=-k_B \int_0^\infty d\omega N(\omega)[f{\rm ln}f
+ (1-f){\rm ln}(1-f)]   ,
\end{equation}
where $f=f(\omega)$  is the Fermi function. Note that the density
of states $N(\omega)$ is calculated self-consistently, using
Eqs. (2.2--2.4).

The presence of resonances at low energies leads to pronounced features
in the low--$T$ specific heat, as shown in  Figure 9. For $T_K\gg T_c$,
$N=2$, these
features are identical to those  predicted by the phenomenological
theory of Refs.\cite{hvw} and \cite{smv}.  For smaller $T_K/T_c$, the
resonance sharpens, as is evident from, e.g., Figure 8.  In this case
resonances in the low-temperature specific heat may be quite dramatic.
Figure 10 shows the temperature dependence of $C/T$
for $N=2$ in the axial state for the case when the resonance is very
narrow, $\Gamma \ll T_c$ and just above the Fermi surface,
$\epsilon_f \ll T_c$.
Such  pronounced features are possible only when
the bare impurity level is close to the Fermi surface,
and the hybridization is weak.
They will be sharper for superconducting states
with larger exponent $r$ in the unperturbed density of states
$N(\omega) \simeq C|\omega|^r$ at low $\omega$.
These anomalies may be observed
experimentally at sufficiently low temperatures.
In this context, it is interesting to note that
a sharp peak in $C/T$ has been observed in the heavy fermion
superconductor $UPt_3$ at 18 mK.\cite{schuberth} This peak is present
at roughly the same position
also in the normal state at magnetic fields $B \ge B_{c2}$,
as might be expected
in a situation where the  Kondo temperature is significantly smaller
than the critical temperature.  If such an interpretation of the measurement
of Schuberth et al. in these terms is correct, we would expect the
size of the peak to scale with other measures of the defect concentration,
such as $T_c$ the size of the specific heat jump, in different samples.
We note, however, that fields of order 1 Tesla would normally
destroy a many-body resonance of the usual magnetic type.

We now discuss the low-temperature response functions of the superconductor
in the presence of Kondo impurities, which have not to our knowledge
been previously calculated in the strongly interacting Fermi liquid regime
of interest.  The effect of Kondo impurities on the electromagnetic
response has been calculated in the phenomenological t-matrix approach
mentioned above, and
used to analyze experiments on heavy fermion
and high--$T_c$ superconductors with impurities, but no microscopic
theory is available.
The London penetration depth is obtained from
$\lambda = {\left [ -{{4\pi}\over c}K(0,0) \right ]}^{-1/2}$ , where
$K(0,0)$ is the electromagnetic response kernel.
The kernel is given by the linear response formula

\begin{equation}
K^{ij}(0,\Omega_m) = - {{3e^2T}\over {2mc}}
\int_{-\infty}^{\infty} d\epsilon \sum_n {\rm Tr}
{\langle {{\hat k}_i {\hat k}_j {\bf G}(k,\omega_n)
{\bf G} (k,\omega_n-\Omega_m) } \rangle }_{\hat k}  .
\end{equation}
In the static limit, $\Omega_m \rightarrow 0$ , the kernel becomes

\begin{equation}
K^{ij}(0,0) = - {{6\pi e^2T}\over {mc}}
\sum_n \int {{d\Omega}\over {4\pi}} {\hat k}_i{\hat k}_j
{{\Delta^2(k)}\over {{(\Delta^2(k)+{{\tilde \omega}_n}^2)^{3/2}}}}  ,
\end{equation}
where the integral represents the angular average.
Here we specialize again to the $N=2$ case. The temperature
dependence of $(\lambda^{11})^{-2}$ in the polar state, which
corresponds to the component of the superfluid density within
the plane containing the line of zeroes of the order parameter,
is shown in Figure 11.  The low-$T$ behavior changes from linear
to quadratic upon doping. A similar result was obtained
earlier\cite{grossetal,muzikar1}
within the phenomenological theories.
The results for $T_k\simeq T_c$
are qualitatively very similar to those shown.
Finally, $\lambda^{-2}$ at $T=0$ along the main axes of symmetry
for the axial and the polar state as a function of impurity
concentration is presented in Figure 12. We note that the largest component
of the penetration depth scales in the case of line nodes as
$\lambda(n)/\lambda_L(0)-1\sim n^{1/2}$ at low concentrations, and
$\lambda^{-2}(n)\sim \log (n_c/n)$ close to the critical concentration.
It is clear that both the concentration
and temperature dependence of the penetration depth components may
be important tests of gap anisotropy.

\section{Conclusions}

We have presented a slave boson theory of
Kondo impurities in superconductors which
has the advantages of being applicable in the Fermi liquid regime
and  being  relatively easy to use in calculating quantities
of experimental interest at all temperatures in the superconducting
state.
The theory
has been shown to reproduce all of the qualitative features of the
physics found by previous theories for large $T_K$, and is asymptotically
in quantitative agreement with "exact" NRG calculations for the
s-wave case.
It is furthermore capable of going considerably beyond currently available
theories in that practical calculations of superconducting
response functions are possible at low temperatures in the superconducting
state, and can be easily generalized to the unconventional states
of great current interest in the heavy fermion and high-temperature
superconductivity problems.

The single exception to this success is the failure to properly describe
of the s-wave superconductor at very large impurity concentrations,
in that the theory predicts an infinite critical concentration.  In
practical terms this is quite academic, as all independent impurity
analyses will break down due to interimpurity interaction effects long
before any putative critical concentration is achieved.  Nevertheless,
this formal shortcoming of the theory exists and must be addressed.
Preliminary analysis of fluctuations about the saddle point
has convinced us that the Gaussian fluctuations
to the scattering amplitude arising from the slave boson dynamics will
be sufficient to induce a critical concentration, in analogy to the
works of Matsuura et al.\cite{min} and Sakurai\cite{sakurai},
and that the theory as it stands
is sufficient to describe the Fermi liquid regime everywhere except
for very large concentrations in the s-wave case.  The difficulty
does not arise in the unconventional
case.

In unconventional superconductors, we have shown that the phenomenological
theories of Refs. \cite{hvw} and \cite{smv} will be reproduced
by the microscopic theory presented here in the limit $N=2$ and
$T_K/T_c\rightarrow\infty$.   The decondensation of the slave boson
amplitude prevents extension of the theory into the high
temperature regime, but qualitatively it is clear that the effect
of lowering the Kondo temperature is to
sharpen the many-body resonance near the Fermi surface, but lower its
weight.  Thus the effect of resonant scattering leading to
low-energy gapless effects in superconducting thermodynamic and transport
properties is reduced.  Impurities with larger orbital degeneracy may
lead to similar resonances away from the Fermi level, possibly similar
to those observed by Maple et al.\cite{maple3} in specific heat
experiments on Pr-doped YBCO.

The effect of Zn and Ni-doping on superconducting properties of YBCO
suggests that these two dopants may be described within
a traditional Kondo-type picture in conjunction with the theory
discussed here.  Ni, with spin one, may possibly be treated
as a higher-degeneracy scatterer  below its Kondo temperature.
As we have seen, such an impurity acts as a weak scatterer compared
to the N=2, $T_K \gg T_c$ resonant scattering case, a possible
model for Zn.  On the other hand,
recent NMR  measurements appear to suggest that a moment forms around the
Zn site.  Within an isolated impurity picture, this would suggest
that $T_K \ll T_c$, where the present theory is not applicable.  A more
plausible explanation, however, is that the local spin correlations
induced by  a missing Cu must be accounted for, as
suggested by Poilblanc et al.\cite{Poil} Further experimental
and theoretical work on this problem is clearly essential.

We wish to thank P. Kumar and K. Ingersent for useful discussions.
Support was received from the University of Florida
Division of Sponsored Research (L.S.B.) and from
the Alexander von Humboldt Foundation (P.J.H.).
We gratefully acknowledge the hospitality of the Institut
f\"ur Theorie der Kondensierten Materie at the University
of Karlsruhe where some of this work was performed.

\newpage

\figure{Dyson equations for the conduction electron and impurity
Green's functions.
\label{fig:1}}

\figure{The critical temperature for an s-wave superconductor
as a function of impurity concentration. The inset shows the slope
of this dependence evaluated at $T_c = T_{c0}$.
\label{fig:2}}

\figure{Conduction electron and impurity spectral functions
in the Kondo limit in an s-wave superconductor for $N=2$ and
$N=4$. The solid and dashed lines correspond to $T_K = T_c$
and $T_K = 10 T_c$, respectively.
The concentration of impurities in all cases is ${\bar n}= 0.4$.
\label{fig:3}}

\figure{The position and spectral weight of the bound states in the gap
for an s-wave superconductor with Kondo impurities.
The solid (dashed) line is the location (spectral weight)
of bound states.
Only one of the bound states is indicated here, the other one
is located at positive energies, symmetrically with respect
to the gap center. Circles (triangles) refer to the position
(spectral weight) obtained from an NRG calculation.\cite{shiba4}
\label{fig:4}}

\figure{Specific heat jump as a function of $T_c/T_{c0}$. We included
the result of the Abrikosov-Gor'kov theory\cite{skalski} for comparison.
\label{fig:5}}

\figure{The derivative of the specific heat jump evaluated at zero impurity
concentration as a function of $T_K/T_c$ for an s-wave superconductor.
The dash-dotted line shows the asymptotic behavior,
$1-1/{\rm ln}(T_K/2T_{c0})$. The dashed line is the asymptotic form
at $T_K/T_c \rightarrow \infty$
found in Refs. \cite{ichinose} and \cite{sakurai2},
$C^* \simeq 1 - 1/{\rm ln}^2(T_K/T_{c0})$.
\label{fig:6}}

\figure{Critical temperature vs. impurity concentration
for unconventional states considered in this work.
The initial slope at $T_{c0}$ is shown in the inset.
\label{fig:7}}

\figure{Conduction electron and impurity spectral functions
in the Kondo limit in a polar state for $N=2$ and $N=4$ .
The solid and dashed lines correspond to $T_K = T_c$ and $T_K = 10 T_c$,
respectively.
The concentration of impurities is ${\bar n}= 0.4$
\label{fig:8}}

\figure{The low-$T$ part of $C/T$ in the polar state
for $N=2$ in the Kondo limit. The inset shows $C/T$ for
$n=0.01n_c$ over the full temperature range.
\label{fig:9}}

\figure{The low-$T$ part of $C/T$ in the axial state.
Note the sharp resonance in the density of states at low
energies (the inset).
\label{fig:10}}

\figure{The inverse square of the penetration depth vs. temperature
at several concentrations of impurities in the polar state.
\label{fig:11}}

\figure{The inverse square of the $T=0$ penetration depth in the two
principal directions as a function of impurity concentration
for the polar (full lines) and the axial state (dashed lines).
\label{fig:12}}

\end{document}